\documentclass{ws-ijmpd}
\usepackage{latexsym,amssymb}

\begin{document}

\markboth{L. Costamante}
{Blazar porperties: an update}

%
\catchline{}{}{}{}{}
%


\title{\vspace*{-1.5cm} 
BLAZAR PROPERTIES: AN UPDATE FROM RECENT RESULTS}

\author{\vspace*{-0.6cm} Luigi Costamante}

\address{W.W. Hansen Experimental Physics Laboratory \&
Kavli Institute for Particle Astrophysics and Cosmology,
Stanford University, Stanford, CA 94305-4085, USA.
luigic@stanford.edu}

\maketitle


\vspace*{-0.5cm}
\begin{abstract}
After a brief critical overview of the main properties of blazars 
and their classification,  some significant results from recent multiwavelength 
observations are summarized, in the context of the jet physics. 
\end{abstract}

\vspace*{-0.2cm}
\section{The 3D space of blazars' main properties}	
Blazars come in many flavours, and can be classified 
according to three main properties of their overall emission:
1) thermal radiation of the nuclear environment, 2) frequency of the two peaks
in the spectral energy distribution (SED), 3) Compton dominance,
i.e. the ratio between the luminosity of the high and low-energy peak.

The first defines the FSRQ and BL Lac classes, according to the presence or absence
of strong, broad emission lines in the optical spectrum, respectively.
To zeroth order, these can be used to estimate the overall energy density of the 
external Optical/UV radiation seen by the relativistic jet. 
This is relevant for determining the dominant emission component by inverse Compton (IC) 
at high energies (Synchrotron Self Compton  or External Compton)  
and for internal $\gamma-\gamma$ absorption. 
However, since the definition of BL Lac is based on ratios between fluxes
(rest frame equivalent width EW$<5\AA$ and Calcium break $<0.4$; i.e. line vs continuum and 
non-thermal continuum vs host galaxy emission, respectively), the classification
is somewhat dependent on the state of the source\cite{landt}. In fact there are 
BL Lacs with line luminosities similar to FSRQ\cite{scarpa} (though on average much lower). 
Further, if the jet luminosity falls 
well below the host galaxy emission, a galaxy might not even be recognized as a BL Lac, 
though hosting a blazar nucleus\cite{landt}.
Nevertheless, there is a clear, physical difference between FSRQ and 
high-peaked BL Lacs (HBL):
HBL do have a much ``cleaner" environment, with upper limits orders of magnitude lower
than for FSRQ on both the line luminosity and the IR emission from the torus.
The latter, if present, would make these sources completely opaque to TeV gamma-rays,
contrary to observations.

The frequency of the synchrotron peak defines the classification in
Low, Intermediate or High-energy peaked objects 
(LBL, IBL and HBL respectively\cite{giommip94}).
Although its precise location  is often difficult to pin down, 
due to insufficient multiwavelength sampling, an excellent proxy 
for the SED properties  is represented by the X-ray spectrum 
(becoming in fact part of the definition):
LBL/FSRQ have X-ray spectra dominated by the IC emission of low-energy electrons, 
and thus present a flat spectral index ($\alpha_x<1$). 
As the peak shifts towards higher energies, the tail of the synchrotron emission starts 
to dominate in the soft X-ray band, 
yielding concave X-ray spectra (the signature of IBL\cite{on231}). 
HBL  are the objects where the X-ray band becomes 
fully dominated by the synchrotron emission of very high energy electrons, 
and present a steep spectrum ($\alpha_x>1$) if due to radiation above the synchrotron peak, 
or a convex/flat spectrum if the peak is inside or above the 
observed X-ray band  (in this case they are called "extreme BL Lacs"\cite{extreme}). 
%
The SED type must be taken into account when comparing the variability properties 
and correlations among wavelengths in different objects.
The key information is the location of the observed band with respect to the SED peaks.
For example, the study of the emission above the peaks
--highly variable and corresponding to the highest energy electrons-- 
requires observations in the X-ray/TeV
bands for HBL, but in the optical/GeV bands for LBL-FSRQ. 
The same Optical/GeV bands, instead, sample the emission before the SED peaks in HBL, 
which is typically much less variable. 
Observations in the X-ray/GeV bands  sample the high/low energy electrons in HBL, 
respectively, while they sample the low/high energy electrons  in LBL-FSRQ
(i.e. the other way round).
 
The Compton dominance determines  the 
apparent bolometric luminosity of the sources, and the ratio between the energy 
densities of the different radiation and magnetic (B) fields seen by the jet.  
The EGRET and VHE observations have shown so far
that FSRQ are on average more Compton-dominated than BL Lacs, showing a sequence of values 
going from 100 in FSRQ to $\lesssim$1  in HBLs.

These three properties seem to correlate, such that blazars form a sequence in this 3D space:
as the luminosity increases, the peak frequency of the two SED humps shifts 
to smaller values (from HBL to FSRQ), and the high-energy peak becomes more 
dominant\cite{gf98}. 
This has been interpreted by Ref\cite{gg98} as due to different amount of cooling
suffered by the electrons: the more severe cooling in FSRQ yield typical electron 
energies smaller than in low-power objects (BL Lacs), and the presence of intense 
external radiation given by BLR photons makes the EC process more dominant for FSRQ.
Recently, this scenario has been further developed connecting the thermal and jet properties
with two more fundamental parameters: 
the black hole mass (M) and accretion rate ($\dot M$)\cite{ggnew}.
Based on the assumption that the size of the dissipation region is controlled by M and
that the BLR exists only if the disk luminosity is above a certain fraction of 
the Eddington luminosity, the average properties of blazars can be recovered with 
the single assumption that the power of the jet $P_{\rm jet} \propto \dot M c^2$.

It should be reminded however that these results are based mostly 
on the (few) EGRET data. The low sensitivity of the detector implies that
the observed properties are biased towards high or flaring states, and depends on the
(unknown) duty cycle of the sources. The {\it Fermi} data will soon provide 
many important answers.

\vspace*{-0.2cm}
\section{The two ends of the electron distribution}
The lowest and highest electron energies ($\gamma_{min}$ and $\gamma_{max}$)
are usually difficult to constrain. 
Synchrotron radiation at $\gamma_{min}$ is emitted in the self-absorbed regime,
while the EC emission
--located in the valley between the SED peaks and in principle observable in the X-ray band--
is often hidden below even a modest SSC emission from higher-energy electrons.
However, in FSRQ the SSC emission becomes strongly suppressed during states with 
high B-field and in presence of strong external radiation. 
In such cases, as shown by 3C454.3 in 2007\cite{gg454}, the low energy cutoff 
can become ``naked" (see Fig. 1), and the X-ray spectrum tells us 
that the electron distribution  extends down to $\gamma_{min}=1$.

At $\gamma_{max}$, the synchrotron emission  is generally covered 
by the onset of the IC emission of lower-energy electrons, for FSRQ/LBL/IBL objects.
In fact, it is often left as free parameter in modelling.
However, its IC emission can emerge at VHE, determining the VHE flux
in a very sensitive way.  
Indeed, the recent detections at VHE of BL Lac\cite{magic} and W Comae\cite{veritas},
while not surprising per se since typically compatible with standard SSC fits, 
show that $\gamma_{max}$ can indeed extend to very high values also in these objects.

\begin{figure}[pt]
\vspace*{-1.2cm}
\centerline{
\psfig{file=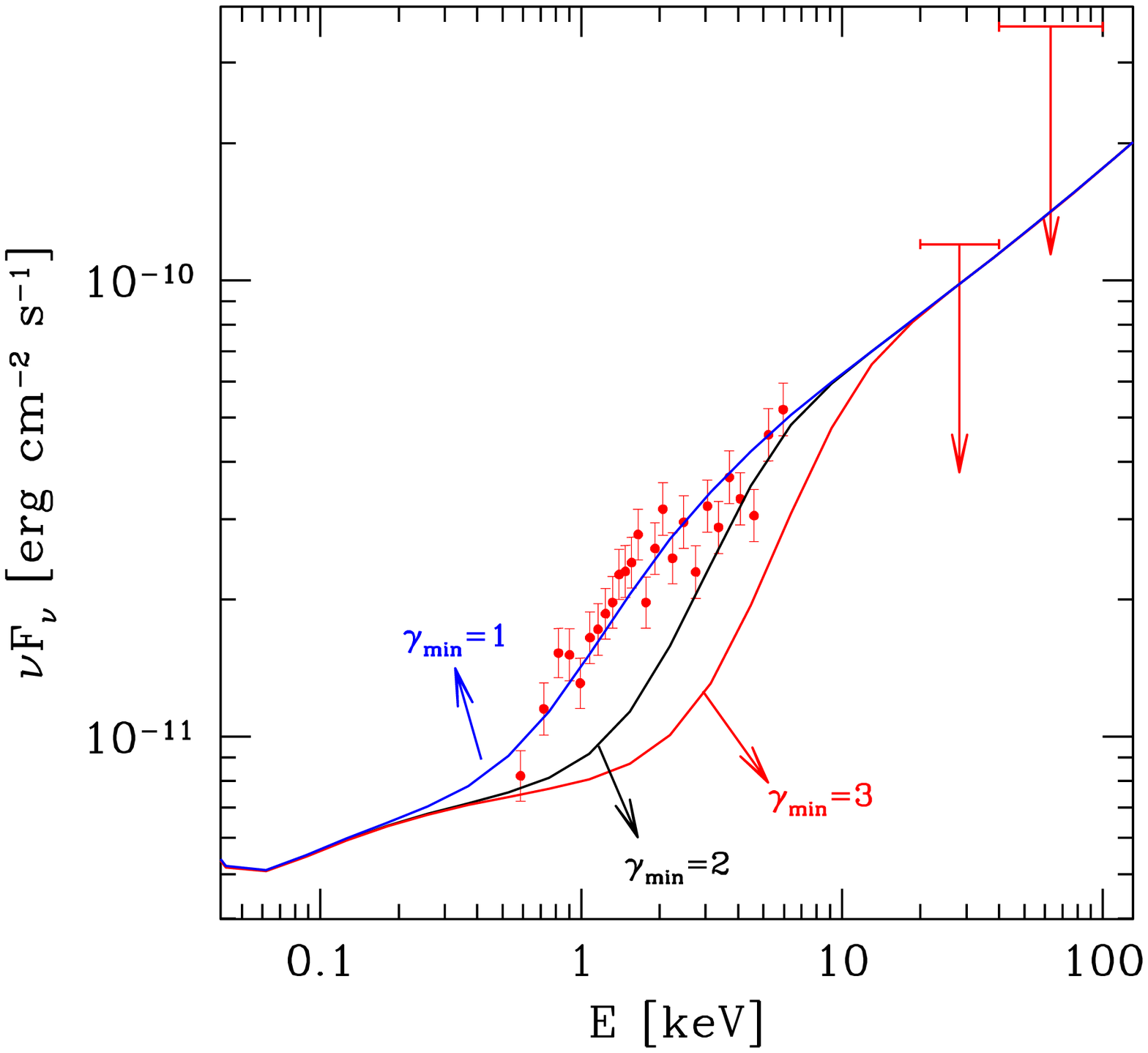,width=3.5cm}
\psfig{file=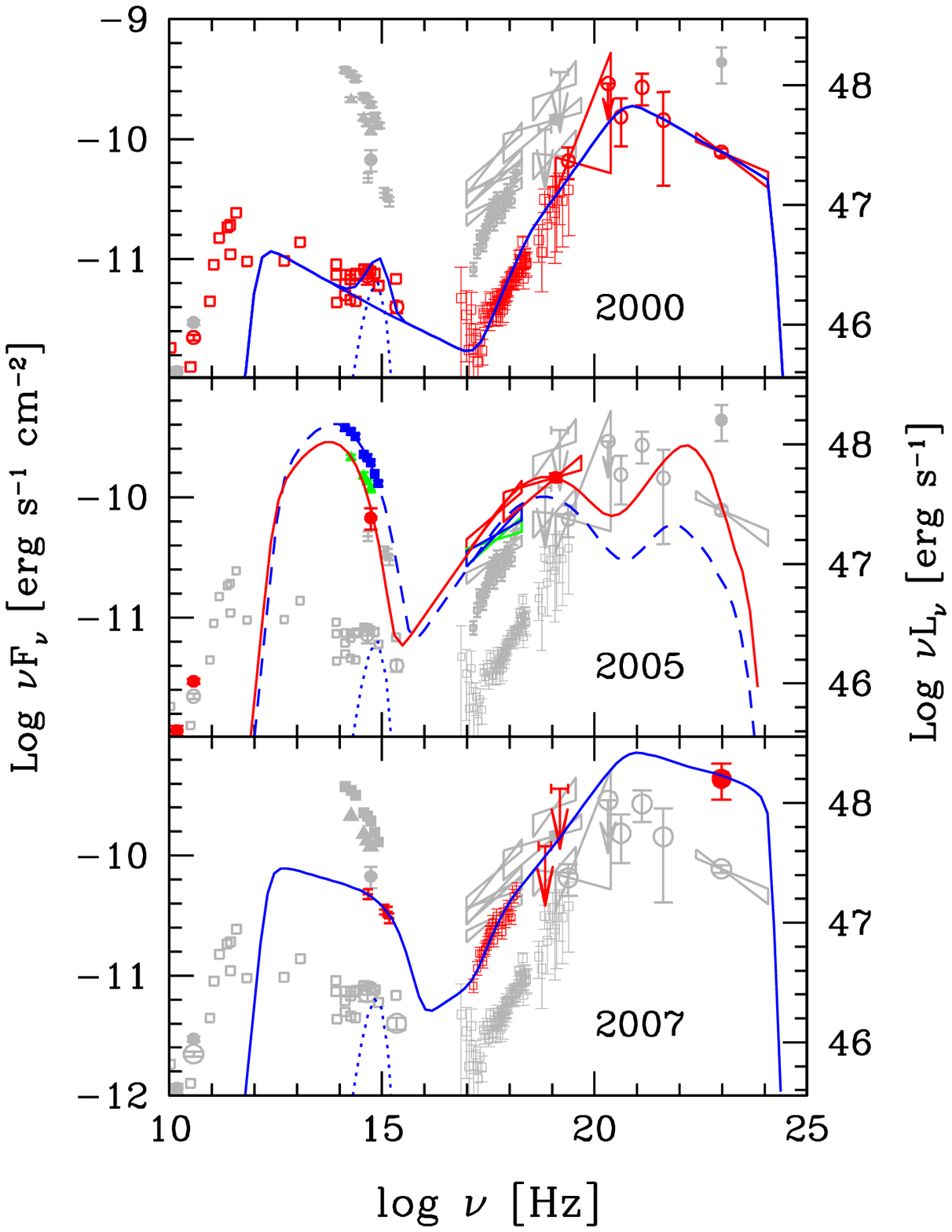,width=4.0cm}
\psfig{file=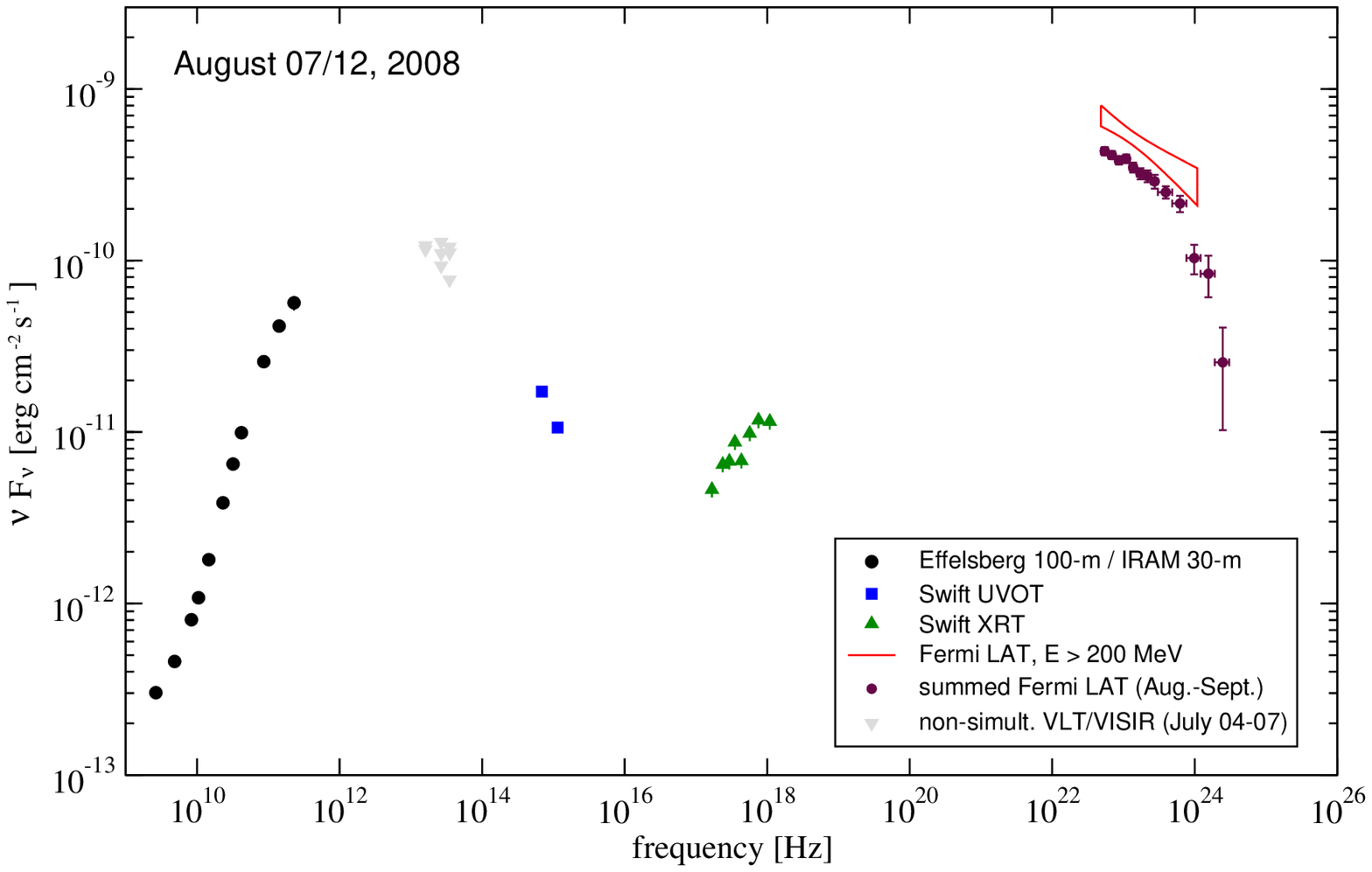,width=5cm}
}
\caption{SEDs of 3C454.3. Left: zoom in the X-ray band. Center: different epochs
(2000-2007). Right: August 2008, first SED with {\it Fermi}-LAT $\gamma$-ray data.
Adapted from Refs$^{8,12}$.}
\end{figure}

\vspace*{-0.2cm}
\section{Location of emitting region: FSRQ vs HBL}
The location of the ``blazar zone" is still highly uncertain. 
Recent data indicate that it might be very different between  FSRQ and HBL.
On the one hand, in FSRQ the  emitting zone cannot be too close to the nuclear regions, 
otherwise the reprocessing of the pairs produced by $\gamma-\gamma$ collisions with the
disk photons would yield X-ray spectra much softer than observed\cite{ggmadau}.
This is also confirmed by the recent multiwavelength campaign on 3C 454.3 (Fig. 1).
In this epoch, {\it Fermi}-LAT also measured a 3.5-day flare 
and for the first time a MeV-GeV spectrum 
with a break around ~2 GeV, likely of intrinsic origin\cite{greg}.
The recent detection of 3C279 at VHE\cite{279} also suggests a location beyond the BLR,
to avoid the otherwise inevitable severe internal absorption on UV photons\cite{hd2155}.
On the other hand, this constrain does not hold for HBL, and 
the very rapid variability observed in PKS\,2155-304\cite{rapid} and Mkn\,501\cite{david} 
as well as M87\cite{m87} 
(corresponding to few Schwarzchild radii) indicates extremely compact regions, 
likely located very close to the black hole.

The recent activity of 3C454.3 is a good example of how the same data 
can leave room to very different scenarios. 
The dramatic SED changes  (Fig. 1) can be explained either
with an accelerating jet of roughly constant power and different locations of the dissipation 
zone inside the BLR\cite{gg454} (closer to the black hole, the blob has lower bulk
motion, higher B and is more compact);  or with a single dissipation zone
located at large distances and variations in jet power and B\cite{sikora}.
In the latter case, at $\sim$10 pc away, the region becomes transparent 
to millimeter wavelengths and the $\gamma$-ray hump is produced 
by EC on IR photons from the dusty torus. 
These two scenarios, however, can be distinguished by specific observations.
Inside the BLR, variability is expected to be shorter due to more compact regions,  
and if EC on UV photons is important, the $\gamma$-ray spectrum should 
exhibit a cut-off at few tens of GeV due to internal $\gamma$-$\gamma$ absorption 
with the BLR photons. Near the millimeter photosphere, instead,
the $\gamma$-ray spectrum can have no sign of cut-off at those energies, 
variability is expected to be not faster than a few days-week and $\gamma$-ray/optical 
flares should be correlated with millimeter outbursts with little or no lag\cite{sikora}.
The {\it Fermi} data will be crucial to this respect.

\vspace*{-0.2cm}
\section{A new mode of flaring for HBL}
In Summer 2006, PKS\,2155-304 was exceptionally active, with
two major $\gamma$-ray outbursts (on the nights of July 27-28 and July 29-30).
During the latter, simultaneous observations performed with {\it Chandra}, HESS and 
the Bronberg optical observatory showed a surprising behaviour\cite{hd2155,2155} (Fig. 2).
On the one hand, the X-ray and $\gamma$-ray emissions were highly correlated,
both in flux and spectrally, with no evidence of lags. 
On the other hand, huge VHE variations ($\sim$22$\times$) were accompanied only by 
small-amplitude X-ray and optical variations (factor 2 and 15\% respectively).
The source showed for the first time in an HBL 
a large Compton dominance  (L$_{\rm C}$/L$_{\rm S}\sim$10) -- rapidly evolving --
and a \emph{cubic} relation between VHE and X-ray flux variations, during a decaying phase.
This behaviour cannot be easily explained in a single-zone scenario\cite{2155}: 
with the observed SED properties 
\begin{figure}[pt]
\vspace*{-0.5cm}
\centerline{\psfig{file=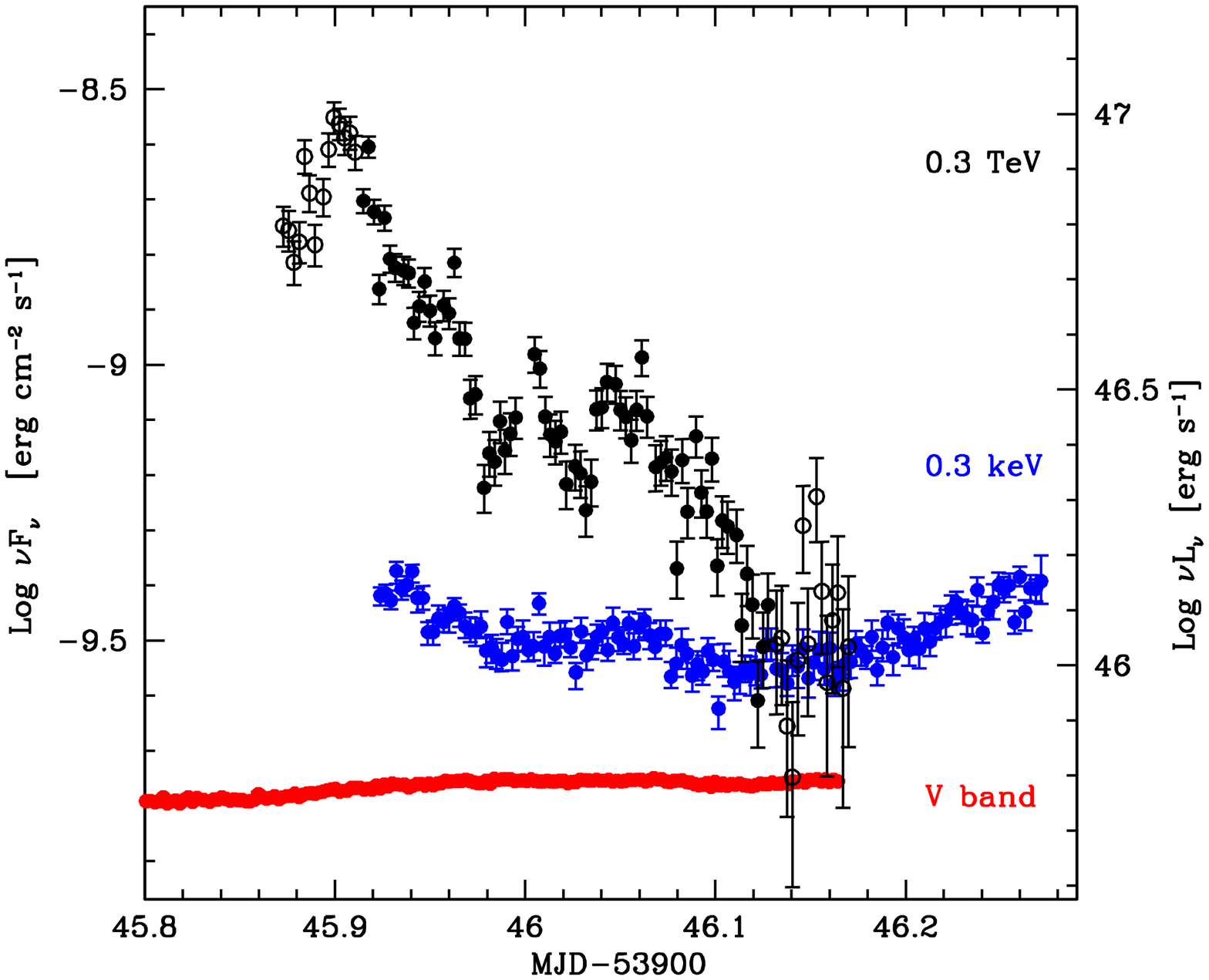,width=4.9cm}
\psfig{file=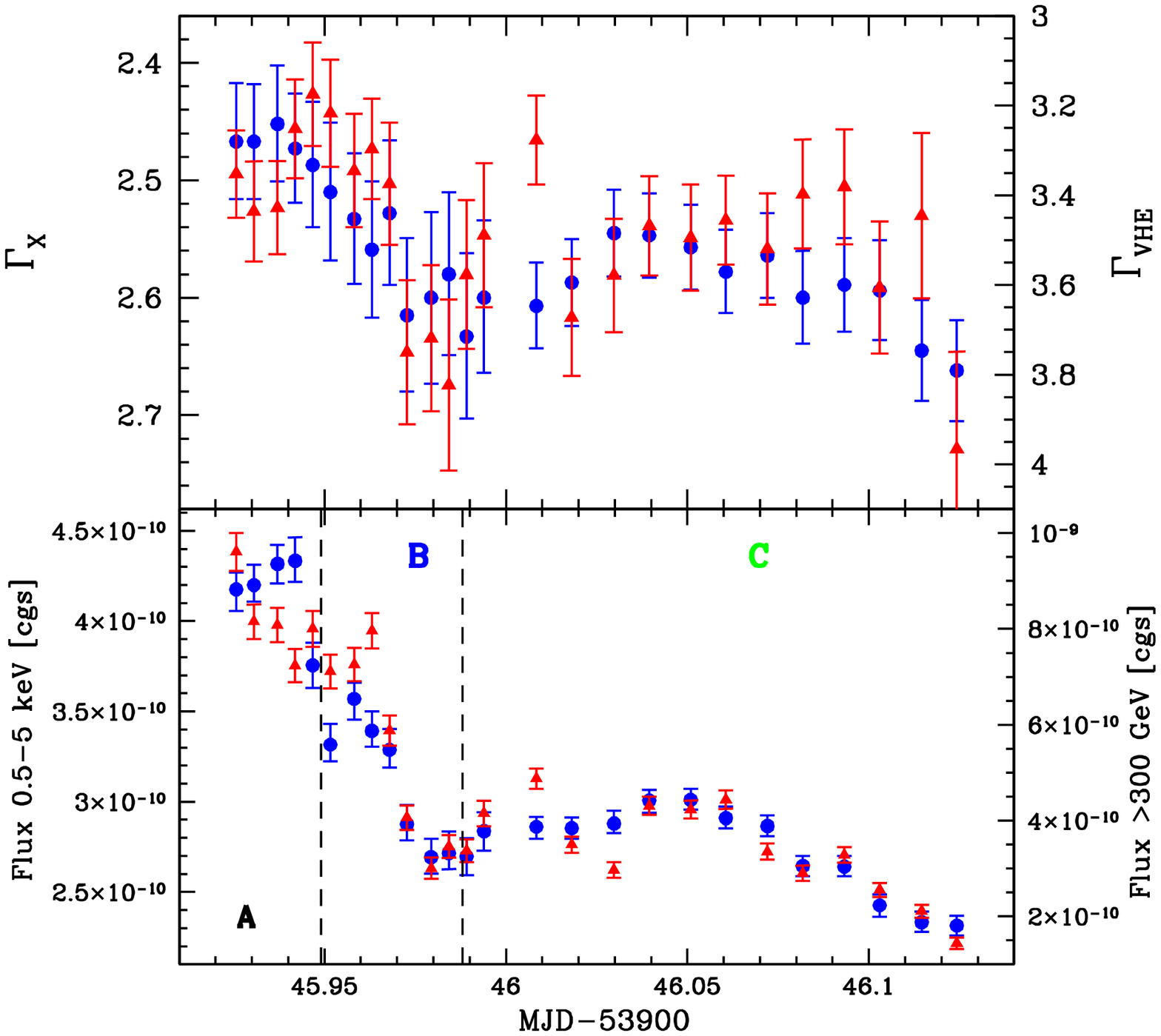,width=4.5cm}
\psfig{file=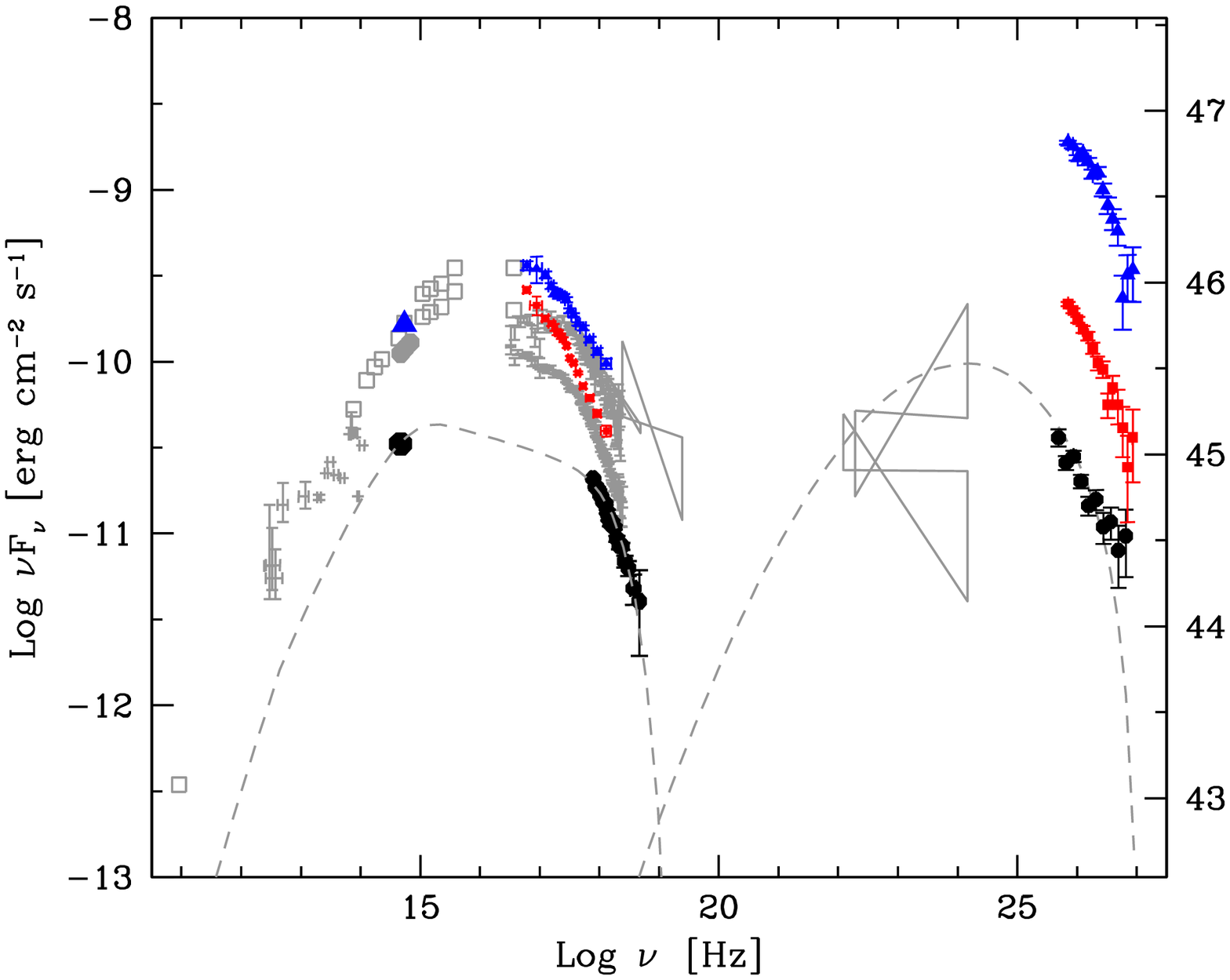,width=5.2cm}
}
\caption{PKS\,2155-304 on 29-30 July 2006. Left: light curves of the $\nu F_{\nu}$ fluxes 
in the 3 bands. Center: evolution of the X-ray (blue circles) and VHE (red triangles) spectra. 
The VHE flux scale  is the cube of the X-ray scale. Right:
intrinsic SED of the high and low states during the night$^{14,19}$. 
}
\vspace*{-0.2cm}
\end{figure}
the decay cannot be due to radiative cooling. Adiabatic cooling due to rapid expansion
could explain most of the features. However, the cubic decay implies that 
B has to increase as the blob expands at a rate $B\propto R^{-m/r}=R^{+0.4}$
(i.e. the magnetic energy has to increase substantially, as $W_B\propto R^{+3.8}$)
and on the same timescales of the VHE variations. This would yield
fast correlated variations in the optical band as well, decreasing by $\sim$15\%.
This is in contrast with the optical lightcurve, which shows a rise
starting simultaneously with the VHE flare but then remaining 
basically constant during the overall X-ray/VHE decay.
A simpler explanation 
is provided by the superposition of two SEDs, produced by two different 
emitting zones: a flaring one and another responsible for the ``persistent", 
historical SED of PKS 2155-304. 
The true variations of the X-ray emission from the flaring zone  
are then seen \emph{diluted} in the ``persistent'' component,
which have comparable or higher synchrotron fluxes.
Such variations instead are fully visible in the VHE band,
because there the contribution of the standard SED is at much lower fluxes
(as measured during the previous 3 years).
A two-zone scenario is rather common for explaining  major flares in blazars.
The main novelty of this event 
is that the bulk of the luminosity of the new component 
is now emitted in the Compton channel instead of the synchrotron channel.
In all previous cases, like Mkn\,421 in 2000,
Mkn\,501 in 1997 
and 1ES\,1959+650 in 2002, 
the Compton luminosity,  even at the flare maximum, was
always equal to or less than the synchrotron power (using the same EBL model).
A bimodality therefore seems to emerge in the mode of flaring for HBL:
either synchrotron dominated or Compton dominated\cite{2155}.

It is intriguing that this difference might simply depend on 
the location of the flaring zone with respect to the region responsible 
for the persistent SED:  if the new flare is taking place far away, 
there is small radiative interplay between the two zones, leading to a typical SSC-type 
flare. When the flare occurs close to it, or close to the black hole where external 
fields are more intense, the outcome are external Compton-dominated flares. 



\end{document}